\documentclass{elsart}

\usepackage{amsmath,amssymb,graphicx,bm}

\newcommand{\be}{\begin{equation}}
\newcommand{\ee}{\end{equation}}
\newcommand{\bea}{\begin{eqnarray}}
\newcommand{\eea}{\end{eqnarray}}
\newcommand{\ba}{\begin{align}}
\newcommand{\ea}{\end{align}}

\newcommand{\Vlk}{V_{{\rm low}\,k}}

\newcommand{\kf}{k_{\text{F}}}

\begin{document}

\begin{frontmatter}

\title{Variational Calculations of Nuclei \\ with Low-Momentum Potentials}
\author{S.K.\ Bogner}
\ead{bogner@mps.ohio-state.edu}
and
\author{R.J.\ Furnstahl}
\ead{furnstahl.1@osu.edu}

\address{Department of Physics,
         The Ohio State University, Columbus, OH\ 43210}

\date{\today}
%

\begin{abstract}
%
Variational calculations of the deuteron and the triton illustrate
that simple wave function ans\"atze become more effective 
after evolving the nucleon-nucleon potential to
lower momentum (``$\Vlk$''). 
This is consistent with many-body wave
functions becoming much less correlated at lower cutoffs, as shown by
two-particle wave functions and pair-distribution functions in nuclear matter.  
These results motivate a program to explore variational many-body calculations of binding
energies and other low-energy nuclear properties using low-momentum
potentials.   

\end{abstract}

\end{frontmatter}


\section{Introduction}
Variational many-body calculations rely on being able to devise
and optimize sufficiently rich wave function ans\"atze.  
The strong short-range repulsion and tensor forces
of conventional nucleon-nucleon potentials
that fit phase-shift data necessitate complicated, 
highly correlated trial wave functions, which limits the effectiveness
of the variational approach.
Even large-scale calculations of light nuclei using
variational Monte Carlo (VMC) achieve accuracy only at
the five  percent level \cite{VMC} 
and Green's function Monte Carlo (GFMC) is needed to
evolve the wave functions \cite{GFMC}. 
Recent results with low-momentum potentials, however, suggest that
more effective variational calculations should be possible \cite{VLOWKMISC}.

The nonperturbative nature of conventional inter-nucleon
interactions can be radically modified
by using the renormalization group to lower the momentum cutoff of a two-nucleon potential
\cite{perturbative}.  For low-momentum interactions with cutoffs around
$2\,\mbox{fm}^{-1}$, the softened potential combined with Pauli
blocking leads to corrections in nuclear matter in the particle-particle
channel that are well converged at second order in the potential.
Calculations of nuclear matter using the low-momentum two-nucleon force
$\Vlk$ with a corresponding leading-order three-nucleon (3N) force from
chiral effective field theory (EFT) exhibit nuclear binding in the
Hartree-Fock approximation, and become much less cutoff dependent upon
adding the dominant second-order
contributions.  At the lower
cutoffs, the iterated tensor interaction in the two-body sector does not
play a major role in nuclear saturation, in contrast to conventional
wisdom.  Thus, the relative importance of
contributions to observables from the tensor force or from three-body
forces is strongly scale or resolution dependent.

Similarly, the correlations in many-body wave functions will exhibit significant
resolution dependence.  The dominance of Hartree-Fock and the onset of
perturbative convergence in the particle-particle channel at smaller cutoffs 
implies that the corresponding
wave functions are much less correlated than those associated with conventional potentials.   
This has the practical consequence that variational calculations should
be effective with much simpler ans\"atze.
In this letter, we illustrate this simplicity through variational
calculations of the deuteron and triton, and by examining the pair-distribution
and two-particle wave functions in nuclear matter at empirical saturation density.  
We are \emph{not} trying at this stage 
to optimize the variational
approach for low-momentum potentials.  Rather our goal is to  
motivate a program to reexamine the application of variational methods to 
binding energies and other low-energy properties of nuclei using low-momentum 
interactions. 

The construction of $\Vlk$ is well documented in
Refs.~\cite{VLOWK,VLOWKFEW}, where it is shown that either
renormalization-group equations or Lee-Suzuki transformations
can be used.  Here we employ the latter, which provide a convenient
formalism to evolve consistent operators beyond the Hamiltonian.
In the notation of Ref.~\cite{Nocore,Navratil}, the evolution of an operator $\widehat{O}$
from a momentum cutoff $\Lambda_0$ to $\Lambda<\Lambda_0$ is given by
\be
\widehat{O}(\Lambda)=\frac{1}{\sqrt{P+\omega^{\dagger}\omega}}(P+\omega^{\dagger})\widehat{O}(\Lambda_0)
(P+\omega)\frac{1}{\sqrt{P+\omega^{\dagger}\omega}},
\label{eq:effoperator}
\ee
where the operator $\omega=Q\omega P$ parameterizes the Lee-Suzuki
transformation, the projection operator $P$ projects onto relative momenta
$k\leq\Lambda$, and $Q$ projects onto $\Lambda<k\leq\Lambda_0$. In the case of the 
evolved Hamiltonian it is convenient
to define $\Vlk(\Lambda)\equiv H(\Lambda)-T$, where $T$ is the ``bare'' kinetic energy operator.
By construction, two-body bound-state properties and phaseshifts are preserved for external relative 
momenta up to the cutoff.
Three- and many-body observables require the consistent addition of a
three-body force to remove cutoff dependence \cite{VLOWKFEW}.

We will show results starting from the Argonne
$v_{18}$ potential \cite{AV18}, since it is used in almost all modern VMC
and GFMC calculations.  However, for cutoffs of $2\,\mbox{fm}^{-1}$ or less,
\emph{all} bare potentials that reproduce nucleon-nucleon phase shifts
up to 350\,MeV lab energy, including EFT potentials at N3LO, 
collapse to the same $\Vlk$ \cite{VLOWK}.  Therefore, the pattern of
results for the full cutoff range shown here does not vary
significantly with different initial potentials.

Since $\Vlk$ is energy independent, 
variational calculations with $\Vlk$ proceed as described in ordinary quantum
mechanics texts (e.g., without special normalizations as needed for
energy-dependent potentials).  
That is, given a trial wave function $\psi_{\rm trial}$, our variational
estimate for the ground state energy at cutoff $\Lambda$ is:
\be
   E_{\rm var}(\Lambda) =
    \frac{\langle \psi_{\rm trial}| T + \Vlk(\Lambda) | \psi_{\rm trial}\rangle}
         {\langle \psi_{\rm trial}|\psi_{\rm trial}\rangle}
         \ ,
\ee
which we minimize with respect to the parameters in $\psi_{\rm trial}$.
Alternatively, we get a variational estimate by diagonalizing 
$T + \Vlk(\Lambda)$ in a truncated basis, where the trial wavefunction
is a linear combination of the basis functions.

\section{Deuteron}

We start with a study of the deuteron binding energy.
The philosophy is that for a weakly bound state one shouldn't need to work
hard, so a simple, generic ansatz should work well.
We  test this as a function of the cutoff.
Our first ansatz is adapted from the discussion of wave
functions in momentum space given long ago by Salpeter
\cite{SALPETER51}, which motivates  
the (unnormalized) $^3S_1$ and $^3D_1$ trial functions for the deuteron
(following the conventions of Ref.~\cite{MACHLEIDT01})
\be
  \psi_0(k) = \frac{1}{(k^2 + \gamma^2)(k^2 + \mu^2)}  \ , 
  \qquad
  \psi_2(k) =  \frac{a\, k^2}{(k^2 + \gamma^2)(k^2 + \nu^2)^2} \ ,
  \label{eq:salpeter}
\ee 
where $\gamma$, $\mu$, $\nu$, and $a$ are variational parameters.  Obviously we
could extend this ansatz in many ways, but our point is to see how well we 
can do without having detailed knowledge about the wave function.
The underlying physics implies that $\mu$ and $\nu$ should 
be roughly the inverse range of the interaction
and $\gamma$ should be close to $(-M_N E_d)^{1/2}$, where
$M_N$ is the mean neutron-proton mass and $E_d \approx
-2.2246\,\mbox{MeV}$ is the deuteron binding energy. Moreover, the 
cutoff in $\Vlk$ implies that the exact deuteron wave function does not
contain high-momentum components. Therefore, the two-body trial wave functions
considered contain the same cutoff on the relative momentum.    

\begin{figure}[t]
\centerline{\includegraphics*[width=3.7in]{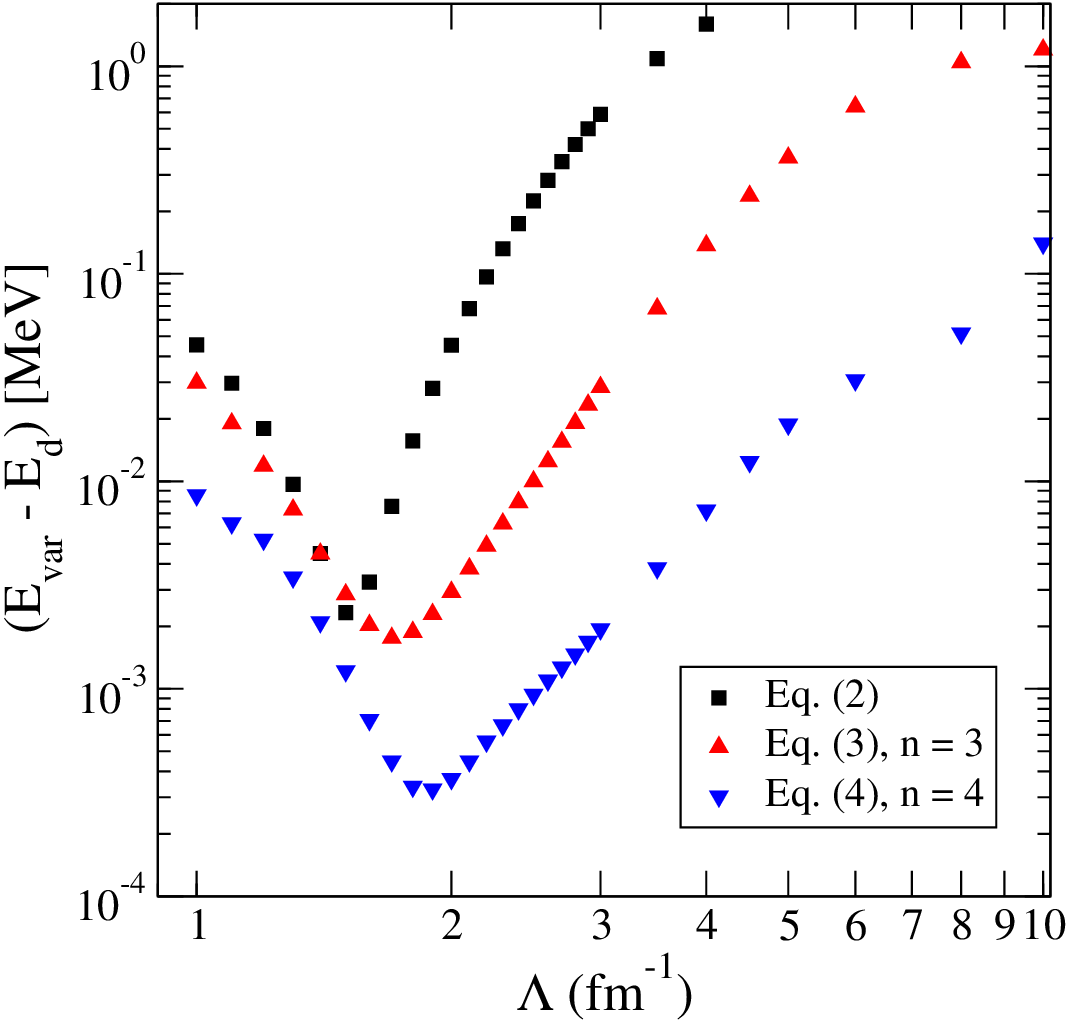}}
\vspace*{-.1in}
\caption{Deviation from $E_d$ of the best variational energy
as a function of cutoff $\Lambda$ for the wave function ans\"atze of
Eqs.~(\ref{eq:salpeter}) and (\ref{eq:machleidt}) 
with different numbers of terms.}
\label{fig:deuteron}
%
%
\end{figure}        

The best variational energy for Eq.~(\ref{eq:salpeter}) 
as a function of the cutoff is shown as the squares
in Fig.~\ref{fig:deuteron}.
These estimates are not even bound for cutoffs above
$\Lambda \approx 5\, \mbox{fm}^{-1}$ (which includes the bare Argonne $v_{18}$
potential) but
rapidly improve as the cutoff is lowered further, reaching
a minimum deviation of less than 3\,keV around $\Lambda
\approx 1.5\,\mbox{fm}^{-1}$.
We emphasize that the low-momentum potential really does preserve
the two-body observables;
directly solving the momentum space
Schr\"odinger equation with 40 gauss points yields the same deuteron binding
energy as the bare potential
to a fraction of an electron volt for all the cutoffs shown.
But variational calculations with the lower cutoffs come much closer
to this energy.

We can also adapt the form used for a high-accuracy representation 
of the deuteron
wave function in Ref.~\cite{MACHLEIDT01} (and elsewhere) to see if the same
pattern holds.
Consider
\be
  \psi_0(k) = \sum_{j=1}^{n} \frac{C_j}{k^2 + m_j^2} \ ,
  \qquad
  \psi_2(k) = \sum_{j=1}^{n} \frac{D_j}{k^2 + m_j^2} \ ,
  \label{eq:machleidt}
\ee
where the $m_j$ are fixed in a geometric progression:
\be 
  m_j = (-M_N E_d)^{1/2} + (j-1)m_0\ , \quad \mbox{with}\ 
  m_0 = 0.9\,\mbox{fm}^{-1} \ ,
\ee
and we treat the $C_j$ and $D_j$ as variational parameters for a given value of
$n$.  
(The very accurate \emph{parameterization} of the deuteron wave function for
the Bonn potential in Ref.~\cite{MACHLEIDT01} has $n = 11$ with some
constraints on the $C_j$'s and $D_j$'s.)
Since the variational coefficients appear linearly, we can simply
diagonalize the Hamiltonian in the truncated basis of Eq.~(\ref{eq:machleidt})
to find the best estimate of the deuteron energy.  
For $n=2$, the best estimate is about 5\,keV above $E_d$
at $\Lambda \approx 1.4\,\mbox{fm}^{-1}$ 
and is unbound above $6\,\mbox{fm}^{-1}$.
By enlarging the basis ($n=3$ and $n=4$ are shown in Fig.~\ref{fig:deuteron})
we find better estimates for all $\Lambda$, but
the steep improvement remains with the minimum shifting gradually
higher.

\begin{figure}[p]
\centerline{\includegraphics*[width=4.1in,angle=0]{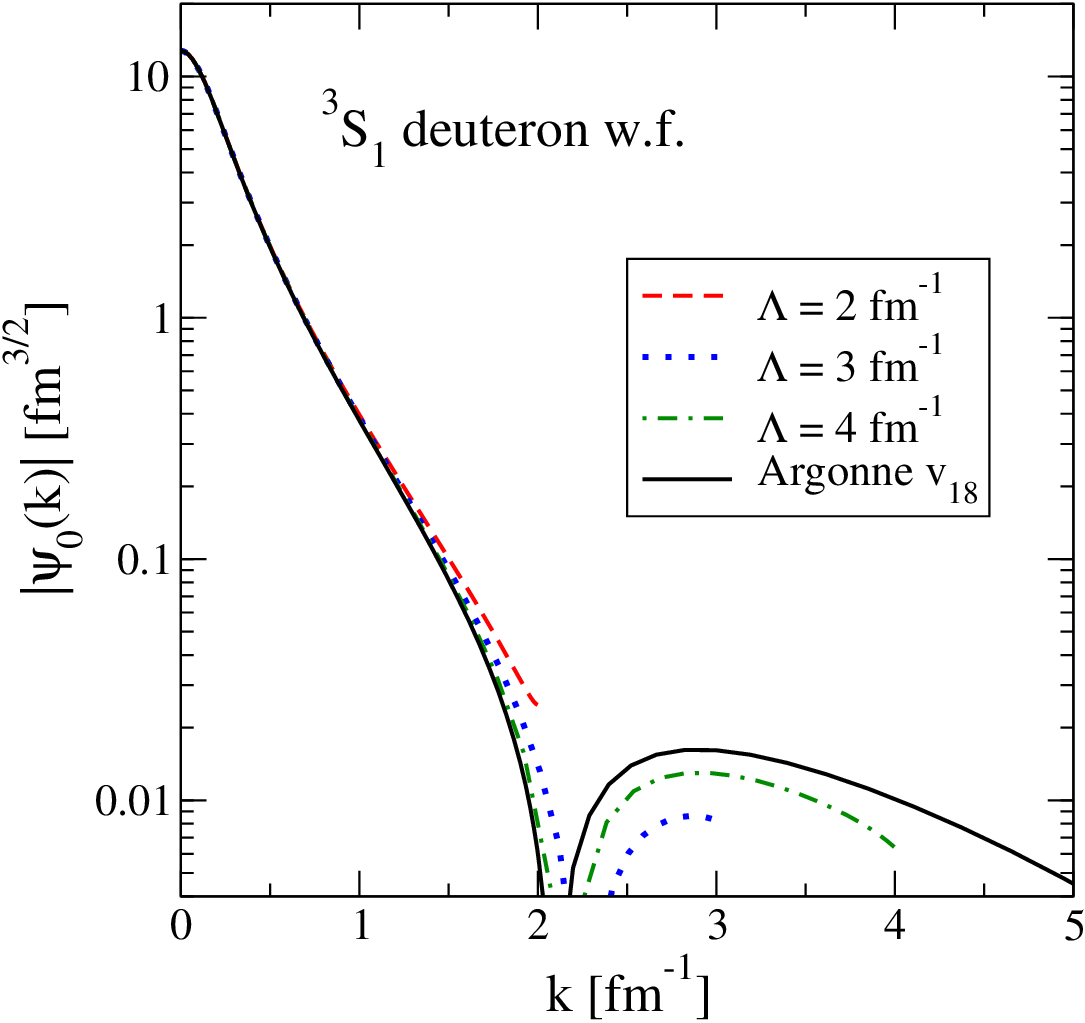}}
\vspace*{-.1in}
\caption{Momentum-space $^3S_1$ deuteron wave function for a range of
    cutoffs and for the bare Argonne $v_{18}$ potential.}
\label{fig:kspacewf}
%
\vspace*{.3in}
%
\centerline{\includegraphics*[width=4.1in,angle=0]{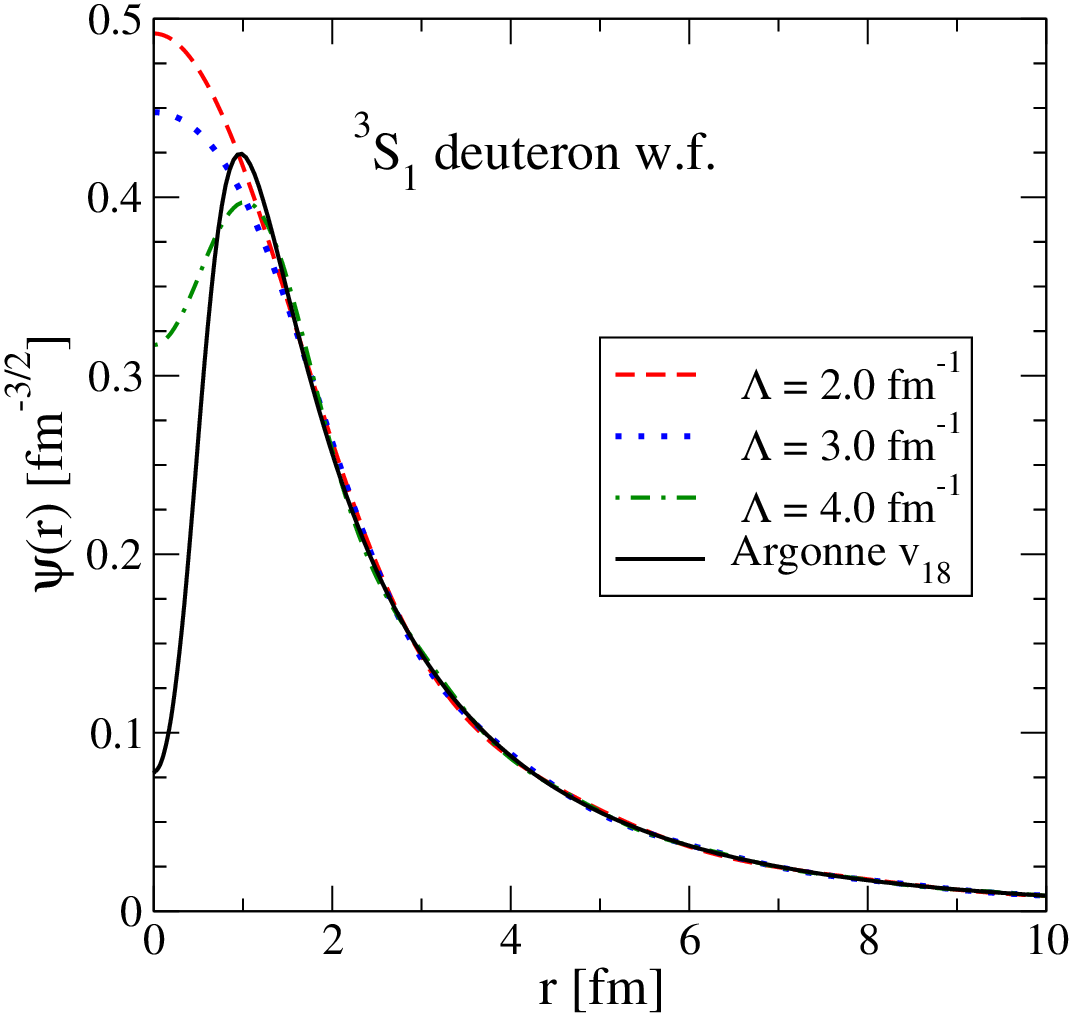}}
\vspace*{-.1in}
\caption{Coordinate-space $^3S_1$ deuteron wave function for a range of cutoffs 
   and for the bare Argonne $v_{18}$ potential.}
\label{fig:rspacewf}
\end{figure}

The success of simple ans\"atze for the deuteron at low cutoffs can
be understood by
looking at the corresponding wave functions.      
In Fig.~\ref{fig:kspacewf}, we show the exact deuteron
wave functions in momentum space for a variety of cutoffs.     
The immediate source of the problem 
for a good energy estimate at higher cutoffs
is the node in momentum space, which
reflects the short-range correlations that are evident as a suppression
of the coordinate-space wave function for 
$r < 1\,\mbox{fm}^{-1}$ (see
Fig.~\ref{fig:rspacewf}) \cite{Garcon:2001sz}.
This short-range/high-momentum behavior is increasingly resolved at
higher cutoffs, which entails finer and finer cancellations in the 
variational integrals.

As suggested by the plots of the $^3S_1$ component $\psi_0(k)$ in
Fig.~\ref{fig:kspacewf} and the analogous behavior of the $^3D_1$
component $\psi_2(k)$, matrix elements of the operator
$a^\dagger_{\bf k} a^{\phantom{\dagger}}_{\bf k}$
(which are proportional to $|\psi_0(k)|^2 + |\psi_2(k)|^2$) change as
$\Lambda$ is lowered.
This is neither surprising nor
worrisome, since the momentum distribution defined by the ``bare''
$a^\dagger_{\bf k} a^{\phantom{\dagger}}_{\bf k}$ operator is not an
observable (see Ref.~\cite{FURNSTAHL02} for a detailed discussion), 
and more generally because operators beyond the Hamiltonian must also be
evolved in order to give $\Lambda$ independent expectation values \cite{HAXTON00}. 
However, as a test of our crude variational wave functions,
it is instructive to start with this operator and a potential at a large
cutoff $\Lambda_0$ (e.g., $10\,\mbox{fm}^{-1}$) and evolve both down to
$\Lambda$ using Eq.~(\ref{eq:effoperator}). 
By construction, matrix elements with the exact wave functions as in
Fig.~\ref{fig:kspacewf} are unchanged even for $k > \Lambda$; we have
verified this explicitly.
One might worry that an intricate interplay of
operator and wavefunction is needed to preserve the matrix element and
this might be lost with a variational wave function, but this is not the case
with low momentum interactions.
For example, matrix elements of the evolved operator with the
simple variational wave function of Eq.~(\ref{eq:salpeter}) for $\Lambda
\approx 1.5\,\mbox{fm}^{-1}$ (which gives the best energy estimate)
underpredict the exact result by less than 4\% for $k <
0.1\,\mbox{fm}^{-1}$ and then closely reproduce it for all $k$ up to
$5\,\mbox{fm}^{-1}$.
This issue will be examined in more generality in future work.


%

\section {Triton}

Moving from the deuteron to the triton is a significant step in complexity, but
we can still test fairly simply whether the basic deuteron results carry over by using
a truncated harmonic oscillator basis for a variational 
calculation. The antisymmetric three-nucleon basis is generated from
the Jacobi coordinate oscillator states~\cite{Navratil}
\be
\mid\!(nlsjt;\mathcal{NL}\frac{1}{2}\mathcal{J}\frac{1}{2})JT\rangle,
\label{eq:jacobi}
\ee 
where $(nlsjt)$ and $(\mathcal{NL}\frac{1}{2}\mathcal{J}\frac{1}{2})$ are the quantum numbers
corresponding to the two relative Jacobi coordinates [e.g., 
${\bf k}=\frac{1}{2}({\bf p}_1-{\bf p}_2)$ and
${\bf q}=\frac{2}{3}({\bf p}_3-\frac{1}{2}({\bf p}_1+{\bf p}_2))$], 
and the basis is
truncated according to the total number of oscillator quanta $N=(2n+l+2\mathcal{N}+\mathcal{L})\le N_{max}$. 
Diagonalizing the intrinsic Hamiltonian in the truncated basis
and minimizing with respect to the oscillator 
length parameter $b$ provides a variational estimate to the true ground-state 
energy.  

\begin{figure}[t]
\centerline{\includegraphics*[width=3.70in]{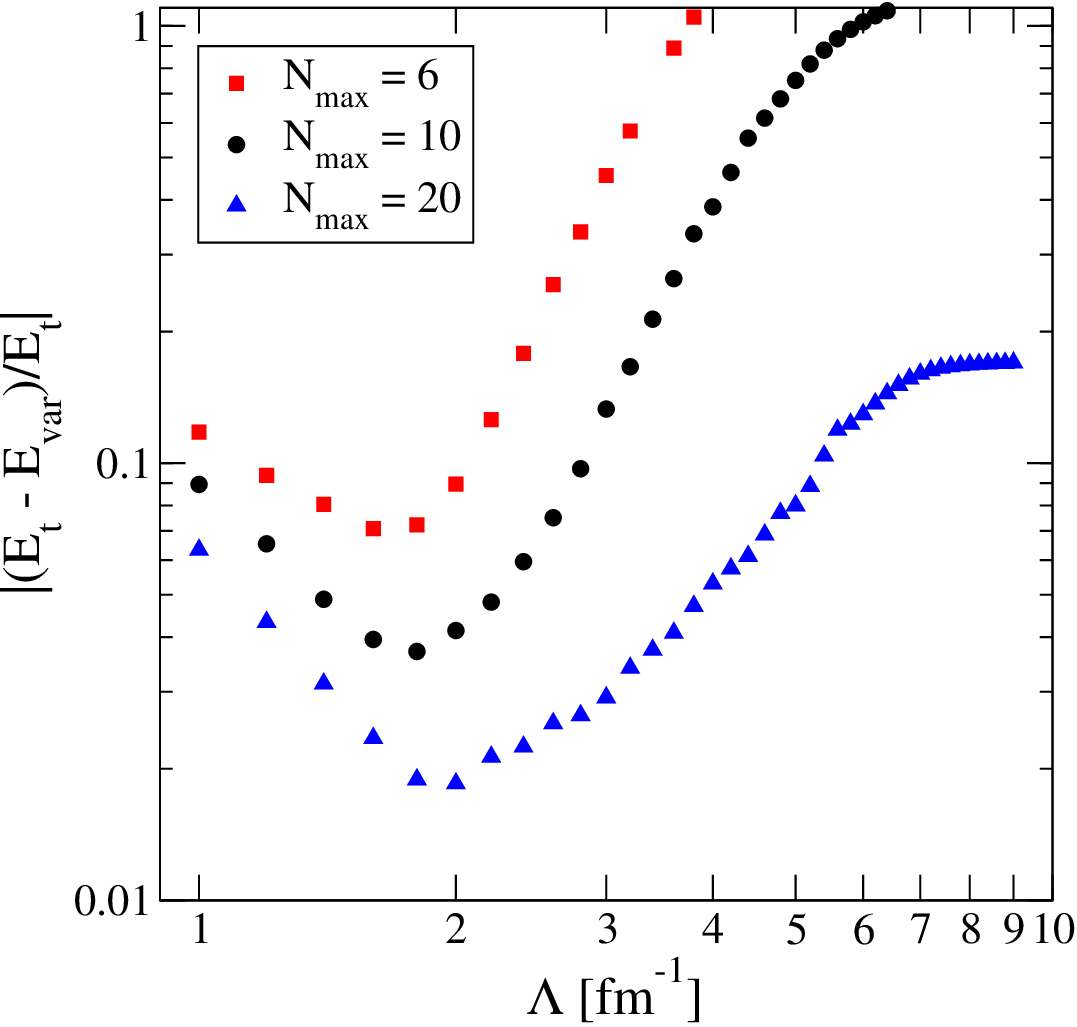}}
\vspace*{-.1in}
\caption{Relative deviation of the variational energy from the Faddeev result
for the triton ($E_t$) with two-body interactions only, as a function of cutoff $\Lambda$
for several truncated oscillator basis sets.}
\label{fig:triton}
\end{figure}

 
Results for the triton with two-body interactions only are shown in
Fig.~\ref{fig:triton}.  Since the three-body contribution varies with the
cutoff, the reference value for two-body alone varies as well.  
Therefore, the results are
given at each $\Lambda$ with respect to the exact 
Faddeev result \cite{VLOWKFEW}
\emph{at that $\Lambda$}, which we label $E_t$.

We see in the figure the same qualitative pattern as found for the deuteron. 
We can calibrate the efficacy of the harmonic oscillator basis at
lower cutoffs by comparison to the calculation by Nunberg et al.\
\cite{HOVAR}, which used the Reid soft-core interaction (which actually has
quite a repulsive core).
Even using a basis with an
additional nonlinear variational parameter compared to the current calculation,
their predicted triton energy is not even negative until
$N_{\rm max} \geq 12$ and the largest calculation with $N_{\rm max} =
28$ yields $E_t = -6.7\,\mbox{MeV}$, which is extrapolated to
$E_t = -7.3\pm 0.2\,\mbox{MeV}$ for $N_{\rm max} \rightarrow \infty$.
That is, $N_{\rm max} = 28$ is still 600\,keV above the converged result.
In contrast, for $\Vlk$ with $\Lambda = 1.8\,\mbox{fm}^{-1}$, the 
result for $N_{\rm max} = 6$ is already only 620 keV above the
converged result, and this drops to 160\,keV for $N_{\rm max} = 20$.\footnote{
We anticipate even better variational convergence
properties if one uses a smooth momentum cutoff in $\Vlk$.} 

\section{Nuclear Matter}

We are optimistic that the general pattern we have seen in simple two- 
and three-nucleon variational calculations will continue for heavier
systems (and with the  inclusion of the  corresponding low-momentum
three-body force), and that the variational  improvement upon  lowering
the cutoff is universal.  Our optimism is based on calculations of
nuclear matter with low-momentum potentials, which are  discussed in
Ref.~\cite{perturbative}. 

Two convenient measures of correlations in  nuclear matter are the
in-medium pair wave functions and the pair-distribution  function.
Working within the Brueckner approximation, the relative in-medium
wave  function for a pair of  nucleons with total momentum $\bf{P}$ and
relative momentum $\bf{k}$ is 
\be 
\mid\!\Psi_{\bf{k}}^{\bf{P}}\rangle=\mid\! {\bf{k}}\rangle+\frac{Q}{\omega-H_{0}}G^{\bf{P}}(
\omega)\mid\!\bf{k}\rangle, 
\ee 
where $Q$ is the Pauli blocking operator and $G$ is the usual Brueckner
$G$ matrix, which sums the in-medium particle-particle ladder diagrams.
In the current calculation, all self-energy effects are neglected for
simplicity, which corresponds  to setting $\omega=k^2+\frac{1}{4}P^2$
and using the pair kinetic energy operator for $H_0$. We stress that
the overall pattern of our results are not changed by this assumption. 
At the same level of approximation, the pair-distribution function is 
given in terms of the coordinate-space pair wave functions \cite{LOWYBROWN}, 
\be
g(r)=\sum_{ST}(2T+1)(2S+1)
     \sum_{\bf{|\frac{P}{2}\pm k|}}^{k_F}\,
     \bigl|\Psi_{\bf{k}}^{\bf{P}}(r;ST)\bigr|^2.
\ee
Physically, $g(r)$ corresponds to the correlation function for finding
another nucleon a distance $r$ from the first.

\begin{figure}[p]
\centerline{\includegraphics*[width=4.in,angle=0]{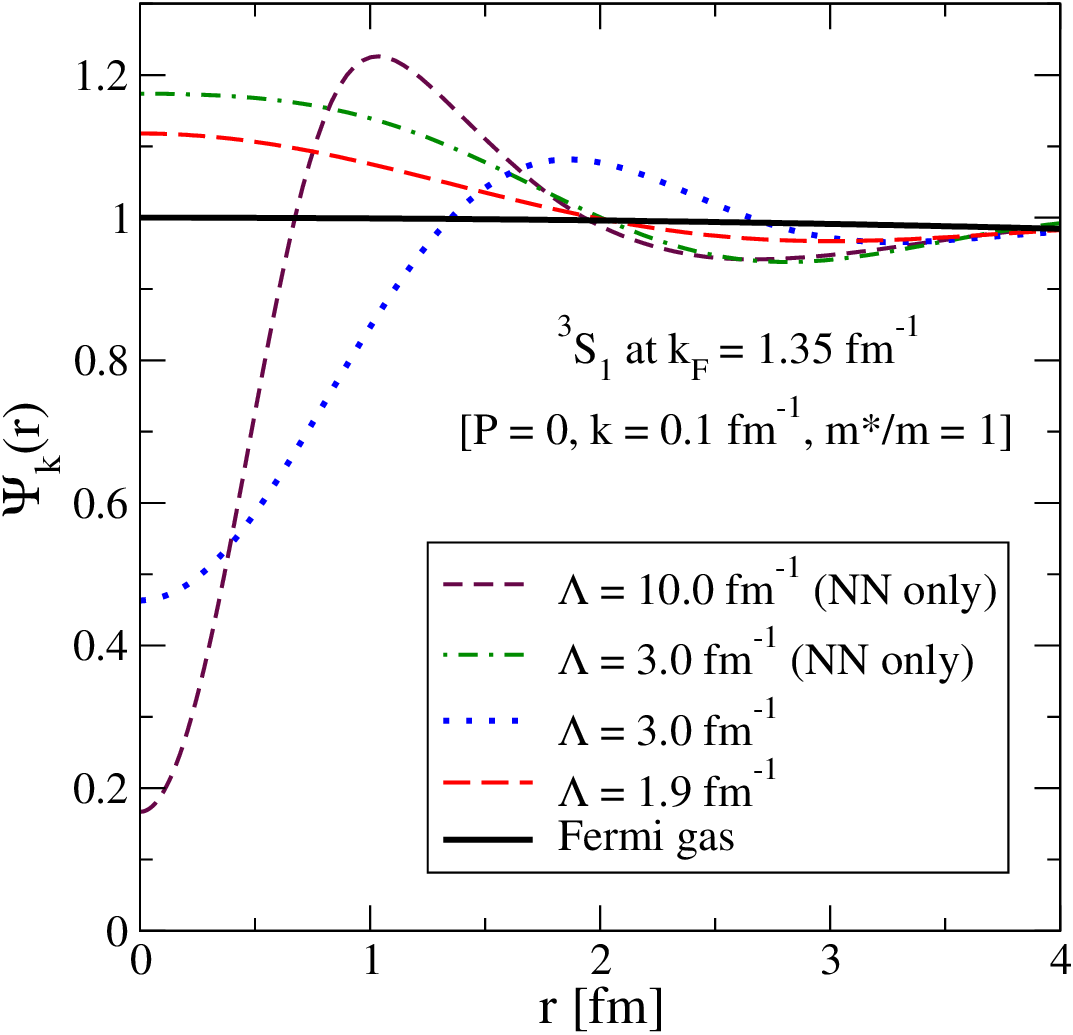} }
\vspace*{-.1in}
\caption{Two-particle wave function for the $^3S_1$ channel
in symmetric nuclear matter at $\kf = 1.35\,\mbox{fm}^{-1}$.}
\label{fig:wf3S1}
%
\vspace*{.3in}
%
\centerline{\includegraphics*[width=4in,angle=0]{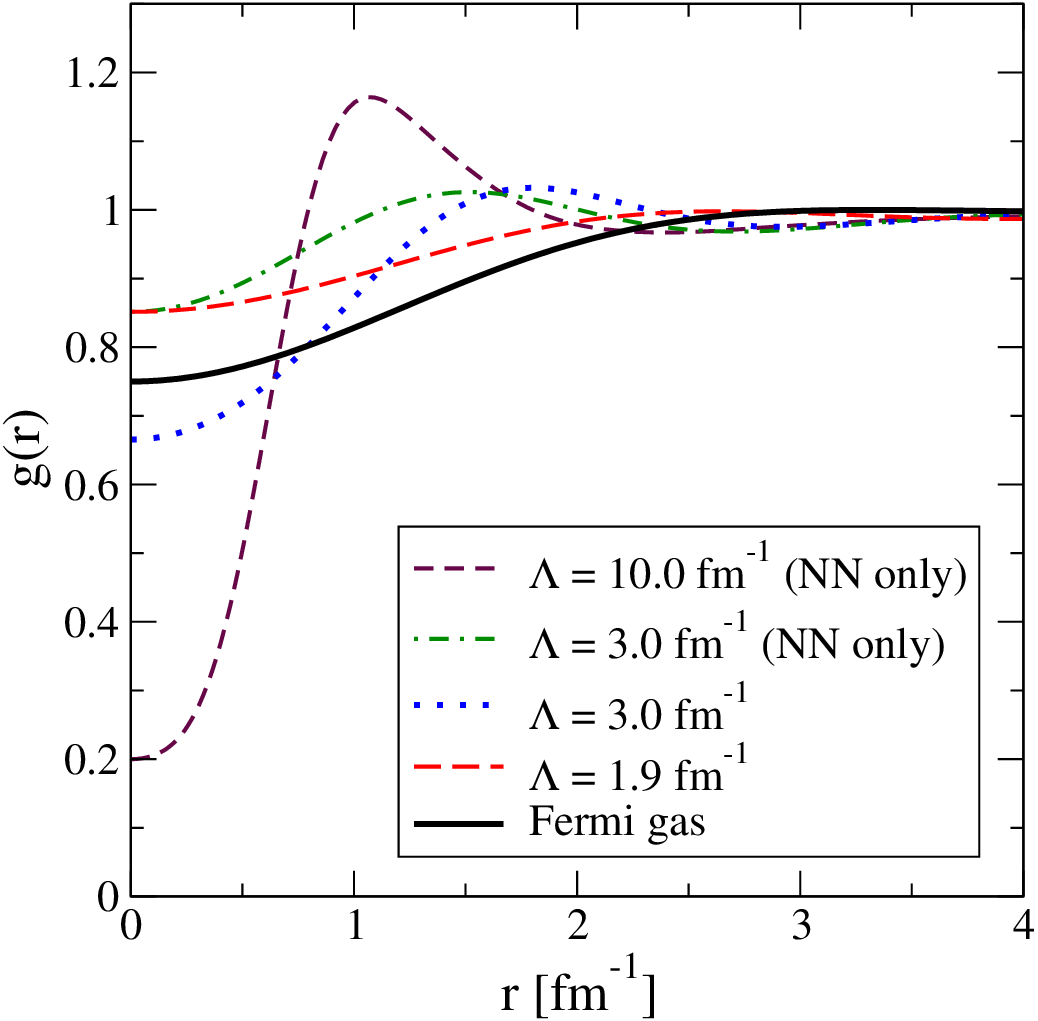}}
\vspace*{-.1in}
\caption{Pair-distribution function in symmetric nuclear matter at $\kf = 1.35\,\mbox{fm}^{-1}$.}
\label{fig:pairdist}
\end{figure}

In Fig.~\ref{fig:wf3S1} we show 
the $^3S_1$ wave function in the
$^3S_1$--$^3D_1$ coupled-channel
for a pair of nucleons with $P=0$ and $k=0.1\,\mbox{fm}^{-1}$, and
in Fig.~\ref{fig:pairdist} the pair-distribution function in 
nuclear matter at empirical saturation density
$\kf = 1.35\,\mbox{fm}^{-1}$. 
Unless otherwise noted, the curves include the dominant effects of the
corresponding three-body force at each cutoff by converting the 3N
vertex into a density-dependent
two-body correction that is added to $\Vlk$, 
see Ref.~\cite{perturbative} for details.

The two-particle wave functions and pair-distribution functions in nuclear matter exhibit the same
promising features we found in the simpler two- and three-nucleon systems. Namely, the strong short-ranged correlations
are ``blurred out'' as the interactions are evolved to lower momenta. This is clear for the two-particle wave function, where
the ``wound'' resulting from the short-range repulsion is more pronounced at the higher cutoffs. Similar results are found for 
the pair-distribution function, where $g(r)$ at smaller cutoffs has little short distance structure and lies fairly close to the
Fermi gas (i.e., Hartree-Fock wave functions) values, where the correlations arise solely from Fermi statistics.  

It is interesting to note that the correlations induced by the three-body force are significantly stronger at
larger cutoffs. At $\Lambda=3.0\,\mbox{fm}^{-1}$, which is the largest cutoff at which the three-body
force was fitted for $\Vlk$ \cite{VLOWKFEW}, the pair-distribution function is suppressed at short
distances (relative to the Fermi gas values) when the three-body force is included. 
In contrast, there is a slight \emph {enhancement} in $g(r)$ at short distances when only the two-body $\Vlk$ is used. 

One finds that the changes in $g(r)$ with and without the three-body force are much less severe at
lower cutoffs. For example, at a cutoff of $\Lambda=1.9\,\mbox{fm}^{-1}$ one finds that $g(r=0)$ decreases by only $0.1$
when the three-body force is included, while the corresponding change for $\Lambda=3.0\,\mbox{fm}^{-1}$ is a factor
of two larger.  This is consistent with the results of
Ref.~\cite{VLOWKFEW}, where it was shown that at smaller cutoffs the three-body force can be accurately treated in perturbation
theory for the triton and alpha particle ground state energies.

As we saw for the deuteron, these plots show that short-range correlations
do not need to be explicit in
the wave functions when calculating low-energy/low-momentum observables
up to nuclear matter densities.  
The key point is that lowering the
resolution makes calculations \emph{simpler}, more efficient, and less
model-dependent. 

\section{Conclusions}  
In summary,
the direct evidence from the deuteron and triton, coupled with the
rapid convergence of the particle-particle channel observed in nuclear matter, 
imply that low-momentum
potentials with $\Lambda \approx 2\,\mbox{fm}^{-1}$
will be much more effective for few-body and many-body 
variational calculations than any conventional large-cutoff potential.
Furthermore, even chiral EFT potentials, which are themselves
low-momentum potentials compared to conventional potentials such
as Bonn, Nijmegen, or Argonne, can be made more effective by
running the cutoff lower. The general idea is to take the EFT
cutoff as large as possible (i.e., in the vicinity of the \emph{breakdown}
scale of the chiral EFT, which presumably is $3\,\mbox{fm}^{-1}$ or higher),
in order to minimize the truncation error \cite{LEPAGE}.  
The evolution to lower cutoffs induces the 
higher-order short-ranged operators that maintain the same truncation error 
in observables as at the higher cutoff. 

To take advantage of these observations,
variational Monte Carlo (VMC) is attractive for its basic simplicity, 
the upper-bound property of the energy estimates, 
and the absence of the fermion sign problem.
Variational calculations with conventional (large cutoff) potentials
are performed in coordinate
space, where the strong correlations are most naturally encoded in
trial wave functions \cite{VMC}.
In fact, most calculations of this sort have used some variety
of the Argonne potential,
which was designed for this purpose and for subsequent GFMC
calculations built on the variational results 
(local and operator based) \cite{VMC,GFMC}.

Based on our results here,
we anticipate more efficient variational
results for low-momentum interactions, with the added advantage of
being able to vary the cutoff as a tool to optimize and probe the quality of the 
solution.
We have no restriction to coordinate space and, indeed, we will first
try to develop
Monte Carlo calculations for light nuclei directly in momentum space. Furthermore,
we can avoid the problem of constructing consistent, model-independent operators
for conventional potentials by evolving to low momentum the potential and operators
from chiral EFT.
More generally,
since Hartree-Fock is a reasonable starting point for many-body calculations, 
the large arsenal of techniques
developed for the Coulomb many-body problem becomes available
and should be explored as well.

\begin{ack}
We thank Achim Schwenk and Thomas Duguet for useful comments and discussions.
This work was supported in part by the National Science Foundation
under Grant No.~PHY--0354916.
\end{ack}


\end{document}